\documentstyle[12pt,a4wide]{article}
\newcommand{\be}{\begin{equation}}
\newcommand{\ee}{\end{equation}}
\newcommand{\bea}{\begin{eqnarray}}
\newcommand{\eea}{\end{eqnarray}}
\newcommand{\bean}{\begin{eqnarray*}}

\newcommand{\eean}{\end{eqnarray*}}
\font\upright=cmu10 scaled\magstep1
\font\sans=cmss10
\newcommand{\ssf}{\sans}
\newcommand{\stroke}{\vrule height8pt width0.4pt depth-0.1pt}
\newcommand{\Z}{\hbox{\upright\rlap{\ssf Z}\kern 2.7pt {\ssf Z}}}

\newcommand{\C}{{\rlap{\rlap{C}\kern 3.8pt\stroke}\phantom{C}}}
\newcommand{\R}{\hbox{\upright\rlap{I}\kern 1.7pt R}}
\newcommand{\CP}{\C{\upright\rlap{I}\kern 1.5pt P}}
\newcommand{\PP}{\hbox{\upright\rlap{I}\kern 1.5pt P}}

\newcommand{\identity}{{\upright\rlap{1}\kern 2.0pt 1}}

\newcommand{\HH}{\mbox{\hbox{\upright\rlap{I}\kern 1.7pt H}}}

\newcommand{\fr}{\frac}

\newcommand{\pr}{\partial}

\newcommand{\dg}{\dagger}

\newcommand{\zb}{{\bar z}}
\newcommand{\pp}{\Delta}
\newcommand{\fac}{(1+\vert z\vert^2)^2}

\font\mybb=msbm10 at 11pt

\def\bb#1{\hbox{\mybb#1}}

\def\bR {\bb{R}}

\def\bC {\bb{C}}

\renewcommand{\CP}{\bC {\rm P}}

\input{epsf}

\newcommand{\news}{\setcounter{equation}{0}}
\begin{document}
\title{\vskip -70pt
\begin{flushright}
\end{flushright}\vskip 50pt
{ \Large  Non-BPS String Junctions and Dyons\\ in
${\cal N}=4$ Super-Yang-Mills}\\[30pt]
\author{Theodora Ioannidou and Paul M. Sutcliffe\\[10pt]
\\{\normalsize  {\sl Institute of Mathematics, University of Kent at Canterbury,}}\\
{\normalsize {\sl Canterbury, CT2 7NZ, U.K.}}\\
{\normalsize{\sl Email : T.Ioannidou@ukc.ac.uk}}\\
{\normalsize{\sl Email : P.M.Sutcliffe@ukc.ac.uk}}\\}}
\date{July 1999}
\maketitle

\begin{abstract}
\noindent 
We construct non-BPS dyon solutions of ${\cal N}=4$ 
$SU(N)$ supersymmetric Yang-Mills theory. These solutions
are the worldvolume solitons which describe non-BPS
Type IIB non-planar string junctions connecting $N$ parallel
 D3-branes. The solutions are smooth deformations of the
$1/4$ BPS states which describe planar string junctions.
\end{abstract}

\newpage
\section{Introduction}
\news\ \ \ \ \ \
Solitons in gauge theories have a geometric interpretation 
as worldvolume solutions describing the intersection of various brane configurations.
For example, the $SU(2)$ BPS monopole describes a D-string stretched between two D3-branes and this
can be seen explicitly by graphing the eigenvalues of the Higgs field over $\bR^3$ \cite{Ha}. 
The worldvolume theory of $N$ parallel D3-branes is ${\cal N}=4$ $U(N)$
 super-Yang-Mills and the $6N$ vacuum expectation values (vevs)
 of the Higgs scalars give the positions of the
D3-branes in the six-dimensional space which is transverse to the worldvolume of the D3-branes \cite{Wi1}.
The $SU(N)$ dyon solutions (an overall $U(1)$ decouples as the centre of mass) describe bound
states of D-strings and fundamental strings stretched between the D3-branes. 

The string spectrum also contains states, known as string junctions, in which three or
more strings meet at a point \cite{Sc,ASY,DM,Se1}. If the string junction is planar,
then it can preserve 1/4 of the underlying supersymmetry and it has
been demonstrated \cite{Be1} that these states correspond to 1/4 BPS dyons in ${\cal N}=4$ $SU(N)$
 super-Yang-Mills, where $N\ge 3.$ Recently classical dyon
solutions have been constructed \cite{HHS1,HHS2,KO,LY} which correspond to these
planar string junctions. Furthermore, it is expected that if $N\ge 4$ then there exist
stable non-BPS states, which correspond to string junctions connecting four
or more D3-branes which are non-planar \cite{BK}. In this paper we  present a class
of static non-BPS dyon solutions which describe such non-planar string junctions.
They are smooth deformations of the
$1/4$ BPS states which describe planar string junctions.

\section{Non-BPS Dyons}
\news\ \ \ \ \ \
The bosonic part of the (3+1)-dimensional ${\cal N}=4$ $SU(N)$
 supersymmetric Yang-Mills Lagrangian is
\be
{\cal L}=\mbox{tr}\{-\frac{1}{2}F_{\mu\nu}F^{\mu\nu}+
\sum_{I=1}^6 D_\mu\Phi^I D^\mu \Phi^I+\frac{1}{2}
\sum_{I,J=1}^6 [\Phi^I,\Phi^J]^2\}
\label{lag}
\ee
where $\Phi^I$, $I=1,..,6$ are the six Higgs scalars
and $D_\mu\Phi^I=\partial_\mu\Phi^I-i[A_\mu,\Phi^I].$
The equations of motion which follow from (\ref{lag})
are
\bea
&D_iD_i\Phi^J=D_0D_0\Phi^J+\sum_{I=1}^6[\Phi^I,[\Phi^I,\Phi^J]]
\label{eq1}\\
&D_iF_{ij}-D_0E_j=\sum_{I=1}^6i[\Phi^I,D_j\Phi^I]\label{eq2}\\
&D_iE_i=\sum_{I=1}^6i[\Phi^I,D_0\Phi^I]\label{eq3}
\eea
where Latin subscripts run over the space indices 1,2,3 and
we recognise the last equation as Gauss' law for the electric
field $E_i=F_{0i}.$

From (\ref{eq1}) it is clear that a consistent reduction can be
made by setting any of the Higgs scalars to zero. If we work
with $M\le 6$ active scalars, ie. $\Phi^{M+1}=\ldots=\Phi^6=0$,
then in the gauge $A_0=-\Phi^1$ the static version of the
above equations reduces to
\be
D_iD_i\Phi^J=\sum_{I=2}^M [\Phi^I,[\Phi^I,\Phi^J]], \ \ 
D_iF_{ij}=\sum_{I=2}^M i[D_j\Phi^I,\Phi^I].
\label{red}
\ee
In this gauge $E_i=D_i\Phi^1$ and $D_0\Phi^I=-i[\Phi^I,\Phi^1]$
 so Gauss' law is equivalent to the first equation in
(\ref{red}) with $J=1.$

From the above it is now clear that the case $M=2$ is very special,
since in this case the equations reduce to
\be
D_iD_i\Phi^2=0, \
D_iF_{ij}=i[D_j\Phi^2,\Phi^2], \
D_iD_i\Phi^1=[\Phi^2,[\Phi^2,\Phi^1]].
\label{2higgs}
\ee
The first two equations are the usual ones for a Yang-Mills theory
with a single Higgs field and hence are solved by any static
 monopole solution of the Bogomolny equation
\be
D_i\Phi^2=-B_i
\label{bog}
\ee
where $B_i=\frac{1}{2}\epsilon_{ijk}F_{jk}$ is the magnetic field.
The final equation in (\ref{2higgs}) is then a linear equation
for $\Phi^1$ in the background of the monopole. Given the Higgs vevs
for $\Phi^1$ and $\Phi^2$ (which give the positions of the 
$N$ D3-branes in a two-dimensional space transverse to the
D3-brane worldvolume) then for each solution of the Bogomolny
equation (\ref{bog}) there is a unique solution to the linear
equation for $\Phi^1$, from which the electric charge of the
dyons can be calculated. These are the 1/4 BPS states which
have been constructed recently \cite{HHS1,HHS2,KO,LY}, following
their predicted existence from string theory as the worldvolume
solitons describing planar string junctions \cite{Be1}.
Due to the fact that the solutions (for fixed Higgs vevs) are parametrized by
Bogomolny monopoles then expressions for the mass and electric
charge of these 1/4 BPS states can be obtained as functions on 
the monopole moduli space \cite{To,HL}.

To describe non-planar string junctions requires at least three
active scalars, in order to assign positions to the D3-branes
which are not co-planar in the transverse space.
With three active scalars, ie. $M=3$, equations (\ref{red})
become
\bea
D_iD_i\Phi^2=[\Phi^3,[\Phi^3,\Phi^2]]\label{3h1}\\
D_iD_i\Phi^3=[\Phi^2,[\Phi^2,\Phi^3]]\label{3h2}\\
D_iF_{ij}=i[D_j\Phi^2,\Phi^2]+i[D_j\Phi^3,\Phi^3]\label{3h3}\\
D_iD_i\Phi^1=[\Phi^2,[\Phi^2,\Phi^1]]+[\Phi^3,[\Phi^3,\Phi^1]]
\label{3h4}.
\eea
The final equation is again a linear equation for
$\Phi^1$ (which determines the electric field) in the
background of the other fields. However, the remaining equations
are more complicated than the usual equations for a Yang-Mills theory with
a single Higgs scalar and are not solved by first order
Bogomolny equations (except for special cases such as $\Phi^3=0$,
when the string junction is again planar and we recover the 
1/4 BPS states). This is to be expected since it is known from 
string theory that non-planar string junctions break all
supersymmetries and hence it is no surprise that we need to 
look for non-BPS configurations which are solutions of the
full second order equations rather than first order Bogomolny
equations. In the next section we shall present some solutions
of equations (\ref{3h1})--(\ref{3h4}), which describe non-planar
string junctions.

\section{The Harmonic Map Ansatz}
\news\ \ \ \ \ \
In this section we shall construct spherically symmetric dyon
solutions of equations (\ref{3h1})--(\ref{3h4}) using the
harmonic map approach introduced in \cite{IS3}.

Coordinates $r,z,\zb$ are used on $\bR^3$, where $r$ is
the radial coordinate and $z$ is the Riemann sphere angular
coordinate given by $z=e^{i\varphi}\tan(\theta/2)$, where
$\theta,\varphi$ are the standard angular coordinates on the two-sphere.
In terms of these coordinates equations (\ref{red}) become
\bea
\sum_{I=2}^M [D_r\Phi^I,\Phi^I]=\frac{i\fac}{2r^2}(D_zF_{r{\zb}}+D_{\zb} F_{rz})\label{ymh1}\\
-i\sum_{I=2}^M\lbrack D_z\Phi^I,\Phi^I]+D_r F_{rz}=
\frac{1}{2r^2}D_z({\fac} F_{z{\zb}})\label{ymh2}\\
\frac{1}{r^2}D_r(r^2D_r\Phi^J)+\frac{\fac}{2r^2}(D_zD_{\zb}\Phi^J+D_{\zb} D_z\Phi^J)
=\sum_{I=2}^M [\Phi^I,[\Phi^I,\Phi^J]]
\label{ymh3}
\eea 
The harmonic map ansatz to obtain $SU(N)$ dyons is a simple generalization of the one 
employed in \cite{IS3} and is given by
\be
\Phi^I=\sum_{j=0}^{N-2} \beta^I_j(P_j-\frac{1}{N}), \ \ \ 
A_z=i\sum_{j=0}^{N-2} \gamma_j[P_j,\partial_z P_j], \ \ \ 
A_r=0.
\label{ansatz}
\ee
Here $\beta^I_j(r),\gamma_j(r)$ are real functions depending only on the radial
coordinate $r$, and $P_j(z,\zb)$ are $N\times N$ hermitian projectors,
that is, $P_j=P_j^\dagger=P_j^2$, which are independent of the radius $r.$
The set of $N-1$ projectors are taken to be orthogonal, so that $P_iP_j=0$
for $i\neq j.$
Note that we are working in a real gauge, so that $A_\zb=A_z^\dagger$.
In (\ref{ansatz}), and for the remainder of the paper, we drop the
summation convention.

The orthogonality of the projectors $P_j$ means that the Higgs fields
$\Phi^I$ are mutually commuting, ie. $[\Phi^I,\Phi^J]=0,$ so they
are simultaneously diagonalizable and this allows the eigenvalues to be interpreted
as the positions of the strings in the transverse space.

The explicit form of the projectors is given as follows.
Let $f$ be the holomorphic vector
\be
f=(f_0,...,f_j,...,f_{N-1})^t, \ \ \mbox{where} \ \ f_j=z^j\sqrt{{N-1}\choose j}
\label{smap}
\ee
and ${N-1}\choose j$ denote the binomial coefficients.
Define the operator $\Delta$, acting on a vector 
$h\in \C^N$ as
\be
\pp h=\pr_z h- \fr{h \,(h^\dg \,\pr_z h)}{|h|^2}
\ee
then $P_j$ is defined as
\be
P_j=\frac{(\pp^j f)(\pp^j f)^\dagger}{\vert \pp^j f\vert^2}.
\ee
The particular form of these projectors corresponds to the requirement that
the associated dyons are spherically symmetric (see \cite{IS3} for more
details). 

As stated above, these projectors form an orthogonal set, and have a number
of other special properties, such as the fact that each of them solves
the harmonic map equation
\be
[P,\partial_z\partial_\zb P]=0
\label{hmap}
\ee
of the two-dimensional \CP$^{N-1}$ sigma model (see \cite{Za} for more details).
Using these properties it can be shown (the analysis follows almost
immediately from that given in \cite{IS3}) that substitution of the ansatz (\ref{ansatz})
into the equations (\ref{ymh1}),(\ref{ymh2}),(\ref{ymh3}) results in a set of
coupled ordinary differential equations for the profile functions
$\beta^I_j(r),\gamma_j(r).$ In fact it is convenient to make a change of
variables to the following linear combinations
\be
\beta^I_j=\sum_{k=j}^{N-2}b^I_k, \ \ \ c_j=1-\gamma_j-\gamma_{j+1},
 \ \ \ \ \mbox{for\ \ } j=0,\ldots,N-2
\label{change}
\ee
where we have defined $\gamma_{N-1}=0.$ 

The magnetic charges, $n_k$, for $k=1,..,N-1$, can be read off from the large $r$
behaviour of the magnetic field
\be
B_i
\sim\frac{\widehat x_i}{2 r^2}G
\ee
where $G$ is in the gauge orbit of
\be
G_0=\mbox{diag}(n_1,
\,n_2-n_1,\,\dots,\,n_{N-1}-n_{N-2},\,-n_{N-1}).
\label{defg0}
\ee
In the case of maximal symmetry breaking, which we shall consider here, they are given
by \cite{IS1}
\be
n_k=k(N-k), \ \ \ \ \ k=1,...,N-1.
\label{charges}
\ee
Similarly, the large $r$ asymptotics of the electric field
\be
E_i
\sim\frac{\widehat x_i}{2 r^2}H
\label{defg}
\ee
allow the electric charges (which classically are real-valued)
to be found from the eigenvalues of $H.$ From our ansatz
\be
H=\sum_{j=0}^{N-2} 2(P_j-\frac{1}{N}) (r^2{\beta^1_j}')\vert_{r=\infty}
\ee
so the electric charges are related to the $1/r$ coefficients of 
$b^1_j$ in a large $r$ expansion.

We shall now restrict to the simplest case in which a non-planar string junction can exist,
that is  $M=3$ active scalars, as described earlier. The lowest rank gauge group we can
consider is $SU(4)$, corresponding to the fact that we require at least four D3-branes
otherwise they will always be co-planar. In this case the equations for the profile functions
are (for $I=1,2,3$)
\bea
(r^2 {b^I_0}')'&=&6c_0^2 b^I_0-4c_1^2 b^I_1\cr
(r^2{b^I_1}')'&=&8c_1^2b^I_1-3c_0^2b^I_0-3c_2^2b^I_2\cr
(r^2{b^I_2}')'&=&6c_2^2b^I_2-4c_1^2b^I_1\cr
r^2c_0''&=&c_0(3c_0^2-2c_1^2-1+r^2(b^2_0)^2+r^2(b^3_0)^2)\cr
r^2c_1''&=&c_1(4c_1^2-\frac{3}{2}c_0^2-\frac{3}{2}c_2^2-1+r^2(b^2_1)^2+r^2(b^3_1)^2)\cr
r^2c_2''&=&c_2(3c_2^2-2c_1^2-1+r^2(b^2_2)^2+r^2(b^3_2)^2).\label{su4}
\eea
The boundary conditions at $r=0$ are that $b^I_j(0)=0$ and $c_j(0)=1$ for all $I,j$,
which ensures that the Higgs fields and gauge potential are regular at the origin.
We shall restrict to the case of maximal symmetry breaking $SU(4)\rightarrow U(1)^3$,
(corresponding to the four D3-branes being distinct in transverse space),
in which case the boundary conditions on $c_j(r)$ at infinity are $c_j(\infty)=0$, for
all $j.$ The remaining free parameters, $b^I_j(\infty)$, required to specify a unique
solution to the set of equations (\ref{su4}), determine the vevs of the Higgs scalars
as follows. Evaluating the ansatz (\ref{ansatz}) along the positive $x_3$-axis, which
corresponds to setting $z=0$, and using the change of variables (\ref{change}), results in
\be
\Phi^I(r)=\frac{1}{4}\mbox{diag}(3b^I_0+2b^I_1+b^I_2,2b^I_1+b^I_2-b^I_0,b^I_2-b^I_0-2b^I_1,
-b^I_0-2b^I_1-3b^I_2)
\label{vevs}
\ee
from which the Higgs vevs can be read off in terms of $b^I_j(\infty).$ 
Writing the components of (\ref{vevs}) as
\be
\Phi^I(r)=\mbox{diag}(\Phi^I_1(r),\Phi^I_2(r),\Phi^I_3(r),\Phi^I_4(r))
\ee
then the positions of the four D3-branes in the three-dimensional transverse space
are given by
\be
(x^4_\alpha,x^5_\alpha,x^6_\alpha)=(\Phi^1_\alpha(\infty),\Phi^2_\alpha(\infty),
\Phi^3_\alpha(\infty)) \ \ \ \mbox{for} \ \ \ \alpha=1,2,3,4.
\label{pos}
\ee
Applying equation (\ref{pos}) for finite values of $r$ gives the positions of the strings
forming the string junction and ending on the D3-branes.

Before considering non-planar string junctions it is perhaps instructive to see how the
$SU(4)$ 1/4 BPS solutions \cite{HHS2,KO}, which describe planar junctions, are obtained
in this formalism. To construct a planar junction we place the D3-branes in the
$(x^4,x^5)$-plane; as an example we shall take the positions (\ref{pos}) to be
\bea
(x^4_1,x^5_1,x^6_1)&=&(14,3,0)\cr
(x^4_2,x^5_2,x^6_2)&=&(-18,1,0)\cr
(x^4_3,x^5_3,x^6_3)&=&(14,-1,0)\cr
(x^4_4,x^5_4,x^6_4)&=&(-10,-3,0).
\label{planar}
\eea
These give the Higgs vevs (via (\ref{pos})) and using the formula (\ref{vevs})
the boundary conditions are found to be
\bea
b^1_0(\infty)=32,\ 
b^1_1(\infty)=-32,\
b^1_2(\infty)=24,\cr
b^2_0(\infty)=b^2_1(\infty)=b^2_2(\infty)=2,\cr
b^3_0(\infty)=b^3_1(\infty)=b^3_2(\infty)=0.
\eea
This is a planar example, and the solution can be obtained explicitly in closed
form. Clearly, $b^3_0=b^3_1=b^3_2=0$, and the functions $c_j,b^2_j$ are those
which solve the Bogomolny equation (see \cite{IS1} for a description of how to
obtain these solutions). In this case
\be
c_0=c_1=c_2=\frac{2r}{\sinh{2r}}, \ \
b^2_0=b^2_1=b^2_2=2\mbox{coth} 2r-\frac{1}{r}.
\ee
The linear equations for $b^1_j$ can then be solved by converting
to a diagonal form using the methods of \cite{HHS2}. Explicitly, the solution is
\bea
b^1_0(r)&=&-b^1_2(-r)=\frac{-3sh(2+8ch^2+shch)+2r(sh+16ch+2shch^2+14ch^3)}{rsh^3}\cr
b^1_1(r)&=&\frac{sh(14+31ch^2)-rch(58+32ch^2)}{rsh^3}
\eea
where $sh=\mbox{sinh}2r$ and $ch=\mbox{cosh}2r.$

Graphing the positions of the strings, using (\ref{pos}), results in the planar
string junction displayed in Figure 1; the squares denote the positions of the 
D3-branes.

For general values of the Higgs vevs, corresponding to non-planar junctions, it is
not expected that the solution of (\ref{su4}) can be found in closed form. However,
it is a relatively simple task to obtain the solution numerically using a gradient
flow algorithm with a finite difference scheme. As a simple example, we take the
 positions of the D3-branes to be
\bea
(x^4_1,x^5_1,x^6_1)&=&(-1,1,0)\cr
(x^4_2,x^5_2,x^6_2)&=&(1,0,1)\cr
(x^4_3,x^5_3,x^6_3)&=&(1,0,-1)\cr
(x^4_4,x^5_4,x^6_4)&=&(-1,-1,0)
\eea
which gives the boundary conditions
\bea
b^1_0(\infty)=-2,\ 
b^1_1(\infty)=0,\
b^1_2(\infty)=2,\cr
b^2_0(\infty)=1,\ 
b^2_1(\infty)=0,\
b^2_2(\infty)=1,\cr
b^3_0(\infty)=-1,\ 
b^3_1(\infty)=2,\
b^3_2(\infty)=-1.
\label{bcnp}
\eea
In fact this example is particularly symmetric, which leads to the reduction
$c_2=c_0, b^1_1=0, b^1_0=-b^1_2, b^2_0=b^2_2, b^3_0=b^3_2.$
The solution satisfying the boundary conditions (\ref{bcnp}) is plotted in Figure 2 
(the individual functions can be identified by their asymptotic values).
In Figure 3 we graph the positions of the strings corresponding to this non-planar
string junction. 

If the string junction is non-planar, but is close to a planar junction, then an
approximate solution can be found as a deformation of a 1/4 BPS state.
As an example, consider the small perturbation of the planar junction (\ref{planar})
to
\bea
(x^4_1,x^5_1,x^6_1)&=&(14,3,\epsilon)\cr
(x^4_2,x^5_2,x^6_2)&=&(-18,1,-\epsilon)\cr
(x^4_3,x^5_3,x^6_3)&=&(14,-1,0)\cr
(x^4_4,x^5_4,x^6_4)&=&(-10,-3,0).
\eea
where $\vert\epsilon\vert\ll 1$ is the constant which describes the small displacement
of the first two D3-branes out of the plane. Equations (\ref{su4}) can then be solved to first
order in $\epsilon.$ The boundary conditions are now modified to
\be
b^3_0(\infty)=2\epsilon, \
b^3_1(\infty)=-\epsilon, \
b^3_2(\infty)=0.
\label{pbc}
\ee
The equations for $c_j$ contain only quadratic terms in $b^3_j$, so the $c_j$ 
solutions are not modified to first order in $\epsilon.$ The equations for
$b^1_j,b^2_j$ do not contain $b^3_j$ so they are also not modified. Thus we
are left with solving the linear equations for $b^3_j$, which are precisely the
same as those already solved to obtain $b^1_j$, the only difference now is
the boundary condition (\ref{pbc}). The solution in this case is
\bea
b^3_0(r)&=&-b^3_2(-r)=\epsilon\frac{6r(2ch^3+2ch+2shch^2+sh)-sh(10ch^2+2+9shch)}{12rsh^3}\cr
b^3_1(r)&=&\epsilon\frac{-2rch(ch^2+2)+sh(2ch^2+1)}{2rsh^3}.
\eea
It can be checked that $\vert b^3_j(r)\vert \le 2\vert\epsilon\vert$
for $j=0,1,2$ and all $r$, hence the assumption that $b^3_j(r)$ is small
for all $r$ if the boundary values are small is a consistent one.
This approximate solution is in good agreement with the one obtained numerically
for small values of $\epsilon.$

As we have seen, for junctions which are close to being planar, approximate solutions
can be obtained by solving linear equations in the background of a BPS monopole
configuration. Thus it should be possible to find
an expression, as a function on the monopole moduli space, for the leading order
 contribution to the mass of deforming a planar junction. Since non-planar
junctions are non-BPS states then they are not protected from quantum corrections
by supersymmetry, so it would perhaps be worthwhile to investigate further the properties
of such states which are perturbations of 1/4 BPS configurations.

\section{Conclusion}
\news\ \ \ \ \ \
We have described the construction of a class of static non-BPS dyon solutions
 of ${\cal N}=4$ $SU(N)$ supersymmetric Yang-Mills theory, whose existence was
predicted by string theory, in which they describe non-planar string junctions
connecting $N$ D3-branes \cite{BK}. The solutions we have presented here are spherically symmetric
and their electric charge is determined by the Higgs vevs. The same situation arises
for 1/4 BPS states describing planar junctions \cite{HHS1,HHS2,KO} and subsequently
more general non-spherical solutions were found \cite{LY} containing parameters
that allow the electric charge to take a range of values. These more general solutions
are required in order to make contact with the quantum theory, since the classical
electric charge takes real values but must be restricted to integer values upon quantization.
It would therefore be useful to find the generalization of our spherical solutions, corresponding
to separating the individual monopoles, though this task may be rather difficult given the
 non-BPS nature of the solutions.

String theory predicts that non-planar string junctions should be stable
configurations, despite the fact that they are non-BPS states \cite{BK}. We therefore expect
that the Yang-Mills dyons we have presented are stable solutions and it would be
interesting to investigate this issue. Although the methods we have used are based
upon those developed to find non-Bogomolny monopoles \cite{IS3}, which are expected to be 
unstable, there is an important difference in the present case in that there are no
BPS states with the given charges and Higgs vevs for the non-BPS dyons to decay to.\\

\section*{Acknowledgements}
\news\ \ \ \ \ \
Many thanks to Dave Tong for useful discussions.
PMS acknowledges the EPSRC for an Advanced Fellowship and the grant GR/L88320.\\


\newpage

\begin{figure}[ht]
\begin{center}
{\epsfxsize=12cm \epsffile{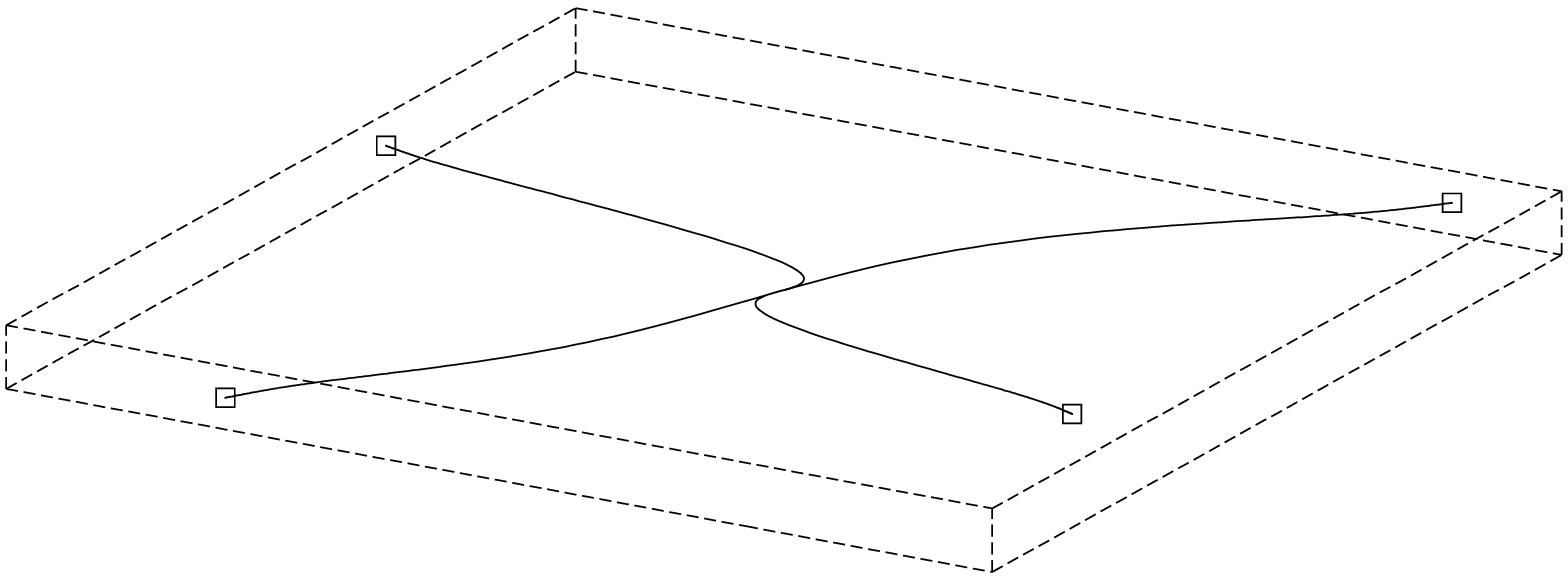}}
\end{center}
\caption{A planar string junction connecting four D3-branes (squares).}
\end{figure}

\newpage

\begin{figure}[ht]
\begin{center}
{\epsfxsize=12cm \epsffile{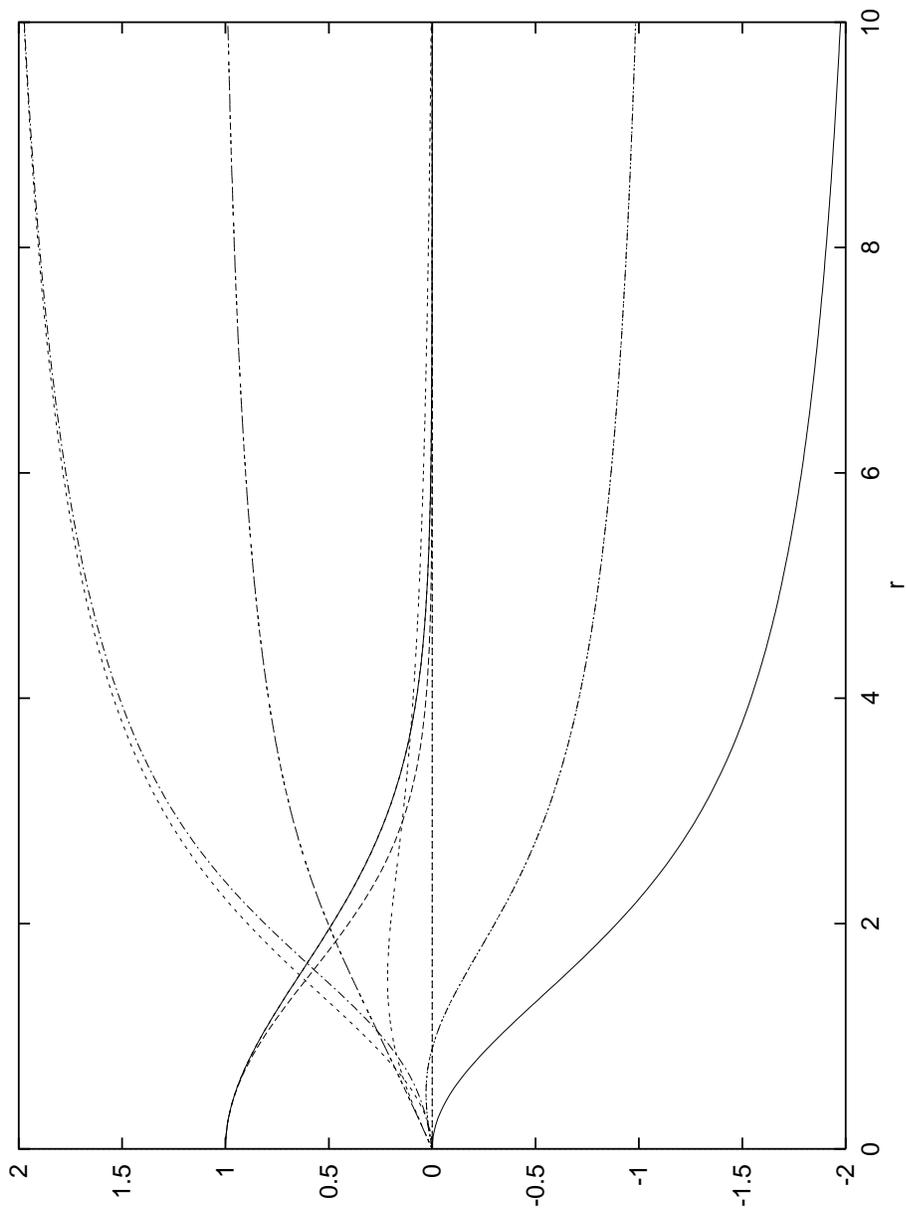}}
\end{center}
\caption{The functions $c_j,b^I_j$ corresponding to the boundary conditions
for the non-planar string junction given in the text. The functions can be 
identified by their asymptotic values.}
\end{figure}

\newpage

\begin{figure}[ht]
\begin{center}
{\epsfxsize=12cm \epsffile{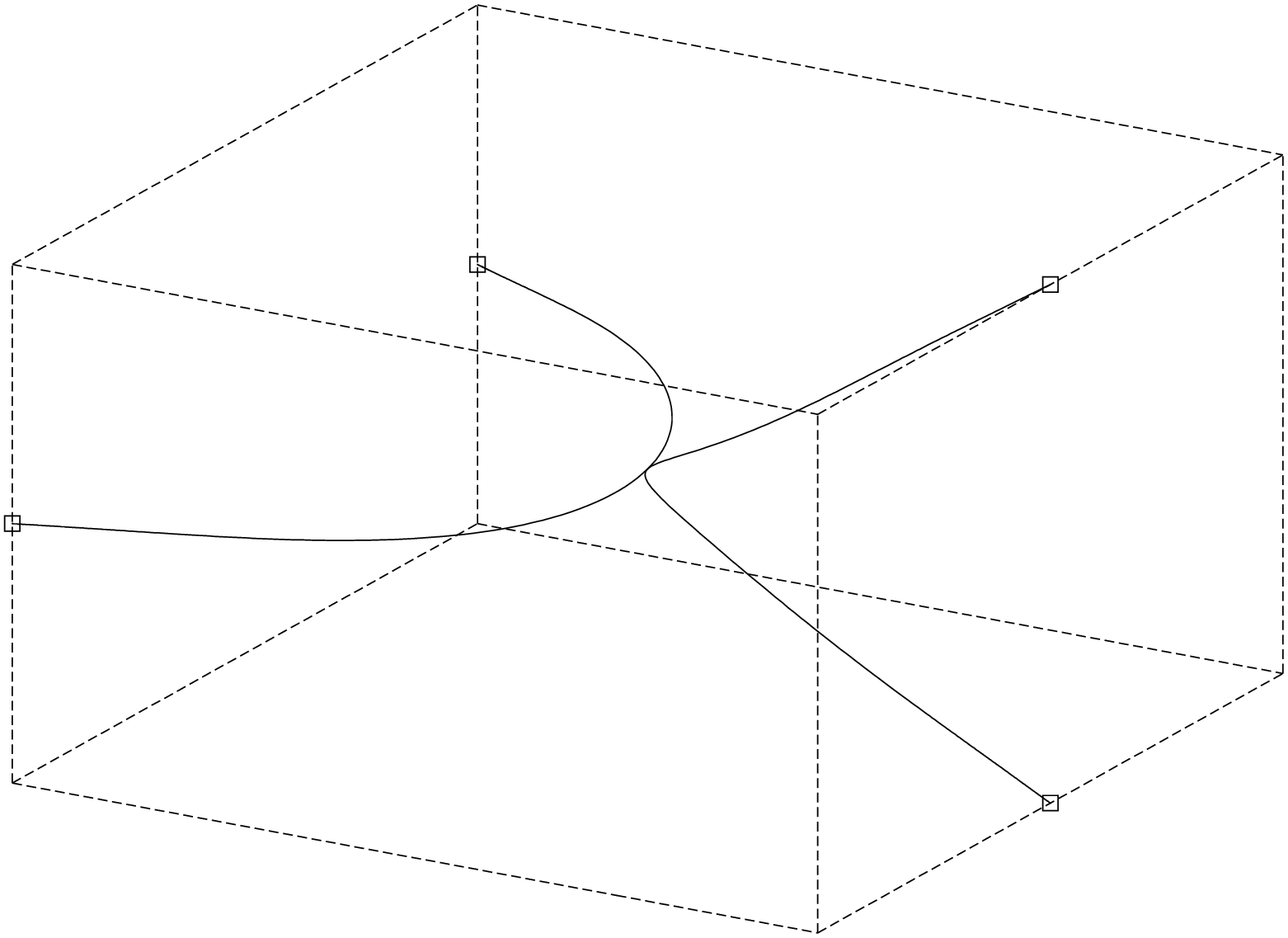}}
\end{center}
\caption{A non-planar string junction connecting four D3-branes (squares).}
\end{figure}


\begin{thebibliography}{99}

\bibitem{ASY} O. Aharony, J. Sonnenschein and S. Yankielowicz, Nucl. Phys. B474, 309 (1996).
\bibitem{Be1} O. Bergman, Nucl. Phys. B525, 104 (1998).
\bibitem{BK} O. Bergman and B. Kol, Nucl. Phys. B536, 149 (1998).
\bibitem{DM} K. Dasgupta and S. Mukhi, Phys. Lett. B423, 261 (1998).
\bibitem{Ha} A. Hashimoto, Phys. Rev. D57, 6441 (1998).
\bibitem{HHS1} K. Hashimoto, H. Hata, N. Sasakura,
 Phys. Lett. B431, 303 (1998).
\bibitem{HHS2} K. Hashimoto, H. Hata, N. Sasakura, 
Nucl. Phys. B535, 83 (1998).
\bibitem{HL} C.J. Houghton and K. Lee, {\sl Nahm data and the mass of 1/4 BPS states}, in
preparation.
\bibitem{IS1} T. Ioannidou and P.M. Sutcliffe, hep-th/9903183 - J. Math. Phys., to appear.
\bibitem{IS3} T. Ioannidou and P.M. Sutcliffe, hep-th/9905169 - Phys. Rev. D., to appear.
\bibitem{KO} T. Kawano and K. Okuyama, Phys. Lett. B432, 338 (1998).
\bibitem{LY} K. Lee and P. Yi, Phys. Rev. D58, 066005 (1998).
\bibitem{Sc} J. Schwarz, Nucl. Phys. Proc. Suppl. 55B, 1 (1997).
\bibitem{Se1} A. Sen, JHEP 03, 005 (1998).
\bibitem{To} D. Tong, hep-th/9902005.
\bibitem{Wi1} E. Witten, Nucl. Phys. B460, 335 (1996).
\bibitem{Za} W. J. Zakrzewski, {\it Low dimensional sigma models} (IOP, 1989).

\end{thebibliography}
\end{document}